\documentclass[pra, showpacs, twocolumn,preprintnumbers, a4paper]{revtex4-1}
\usepackage{graphicx}
\usepackage{amsmath, amsfonts, amssymb, bm}

\usepackage{amsmath}
\usepackage{verbatim}
\usepackage{bm}

\usepackage{color} 

\begin{document}

\newcommand{\ket}[1]{| #1 \rangle} 
\newcommand{\bra}[1]{\langle #1 |} 
\newcommand{\braket}[2]{\langle #1 | #2 \rangle}
\newcommand{\kett}[1]{\left| #1 \right\rangle}
\newcommand{\braa}[1]{\left\langle #1 \right|}
\newcommand{\half}{\mbox{$\textstyle \frac{1}{2}$}}
\newcommand{\beq}{\begin{equation}}
\newcommand{\eeq}{\end{equation}}
\newcommand{\beqa}{\begin{eqnarray}}
\newcommand{\eeqa}{\end{eqnarray}}
\newcommand{\proj}[1]{\ket{#1}\bra{#1}}

\title{Correlations in optically-controlled quantum emitters}

\author{Cristian E. Susa}
\email{cristian.susa.q@correounivalle.edu.co}
\author{John H. Reina}
\email{jhreina@univalle.edu.co}
\affiliation{Departamento de F\'isica, Universidad del Valle, A.A. 25360, Cali, Colombia
}

\date{\today}

\begin{abstract}
We address the problem of optically controlling and quantifying the dissipative dynamics of quantum and classical  correlations in a set-up of  individual quantum emitters under external laser excitation.  We show that both types of correlations, the former measured by the quantum discord, are present in the system's evolution even though the emitters may exhibit  an  early stage disentanglement.   In the absence of external laser pumping, 
 we  demonstrate analytically, for a  set of suitable initial states,  that there is an entropy bound for which quantum discord and entanglement of the emitters are always greater than  classical correlations, thus  disproving an early conjecture that classical correlations are  greater than quantum correlations. Furthermore, we show that quantum correlations can also  be  greater than classical correlations when the system is driven by a laser field. For scenarios  where the emitters'  quantum correlations are below their classical counterparts, an optimization of the evolution of the quantum correlations can be carried out by appropriately tailoring  the amplitude of the laser field and the emitters' dipole-dipole  interaction. We  stress the importance of using the entanglement of formation, rather than the concurrence, as the entanglement measure, since the latter can grow beyond the total correlations and thus give incorrect results on the actual  system's degree of entanglement. 
\end{abstract}



\maketitle

\section{introduction}
The nature and the optimal dynamical control of the correlations present in a given interacting quantum system are of interest to  fundamental  aspects of quantum physics but also to the physical implementation of quantum computation and algorithms \cite{bennett,ladd, knill}.
A decade ago, two separate works reported from Los Alamos \cite{Zurek} and Oxford \cite{Vedral} set a formal framework that allowed the identification of quantum correlations different to entanglement. Such correlations ubiquituosly  manifest themselves in interacting quantum systems, even in the absence of entanglement, and 
 it has been demonstrated that they can play a key role in quantum computing protocols such as DQC1 \cite{dqc1,datta}, without the need to invoke quantum entanglement.  It has become clear that  zero entanglement does not imply classicality, and the task of classifying the kind of total correlations present in  a  given 
system into quantum and classical parts has led to the introduction of quantifiers (different to entanglement) such as the quantum discord (QD) \cite{Zurek}. This  measure of the `quantumness'
of correlations is defined  in terms of the difference between the  two possible ways to obtain the mutual information 
(MI) that arises in the quantum case.  

More recently, 
it has been argued that the power of the quantum computer can be  attributed to both entanglement and quantum discord which are distributed following a monogamic relation \cite{fanchini}.
Further, other reports on the classification  and quantification of quantum and classical correlations have been put forward
\cite{luo1,luo2,ferraro,mazzola,kavan,davidovich}. 
A unified view of quantum, classical and total correlations using 
the concept of relative entropy as a distance measure of correlations has been discussed in \cite{kavan}. An experimental investigation of classical 
and quantum correlations has been recently reported  for a system of two polarized photons in the  presence of a phase-decoherence 
non-dissipative channel \cite{exp1}, and  a sudden transition between classical and quantum correlations dynamics has been reported  \cite{exp1}, a result in line with a recent theoretical  proposal  \cite{mazzola}. 

We start with a brief definition of correlations (Sec. \ref{codef})  followed by the  formulation of the 
dissipative dynamics under consideration, in Sec. \ref{secsystem}. Our main findings are detailed in Sections 
\ref{secdissipative}-\ref{seclaser} as follows: We show that the entanglement of formation, 
rather than the concurrence, is the free of ambiguity metric that should be used to correctly 
quantify the entanglement generated 
in the emitters' system (Sec. \ref{secdissipative}). The computed concurrence can reach higher values than 
the mutual information, which is the highest possible value for the total (classical plus quantum) correlations. In Sec. \ref{secorder} we demonstrate analytically, in the absence of  
laser pumping, that there exist an entropy bound for which the quantum correlations are {\it always} 
greater than classical correlations, a fact which disproves an early conjecture posed in \cite{lindblad}. In Section \ref{seclaser} we give numerical examples for which such result also holds when the emitters are optically-controlled  by an external laser field. This also gives a method  for stationary enhancement of the degree of correlations by tailoring the laser amplitudes and the strength of the 
dipole-dipole interaction.

\section{Quantifying classical and quantum correlations}
\label{codef}

We begin by considering an open bipartite system, say the molecular dimer $AB$. Its quantum dynamics can be described in terms of the total density matrix $\rho_{AB}$; the partial information  about each of the subsystems is given by the reduced density matrices  $\rho_{i}=tr_j(\rho_{AB})$, $i,j=A,B$. 

\subsection{Mutual Information}

In classical information theory, the mutual information  (MI)  can be written in terms of the conditional entropy of two random variables, that is, 
for a system described by two classical random variables, the most information that one can 
gain from the two variables is the MI. 
Let $X$ and $Y$ be two random variables with Shannon entropies $H(X)=-\sum_{x}p_{(X=x)}\text{log}\, p_{(X=x)}$, and $H(Y)=-\sum_{y}p_{(Y=y)}\text{log}\, p_{(Y=y)}$. The MI is 
written as:
\begin{equation}
   I(X:Y)=H(X)+H(Y)-H(X,Y) .
\label{CMI}
\end{equation}
Using Bayes's rules for the conditional probability 
$p_{(X|Y=y)}=p_{(X,Y=y)}/p_{(Y=y)}$,  the MI reads 
\begin{equation}
  J(X:Y)=H(X)-H(X|Y) ,
\label{MIC}
\end{equation}
where $H(X|Y)$ is the conditional entropy \cite{man}.

\subsection{Total correlations}
\label{secTC}


The mutual information has been demonstrated to describe the whole content of correlations in a given  quantum 
system, and at any time $t$, the total correlations present in such a system can be quantified  by the quantum version of
Eq. (\ref{CMI}), the so-called quantum mutual information
\begin{equation}
  I(\rho_{AB})=S(\rho_A)+S(\rho_B)-S(\rho_{AB}) ,
  \label{MI}
\end{equation}
where $S(\rho)=-tr(\rho\text{log}_2\rho)$ is the von Neumann entropy. The 
problem that arises  in the quantum case is that the non trivial version of  Eq.  
\eqref{MIC} is not always identical to $I(\rho_{AB})$. This is because,  
quantum mechanically, 
the conditional entropy implies that the state of a subsystem (say $A$) 
is only known after a set of measures has been carried out on the other 
subsystem ($B$). Ollivier and Zurek \cite{Zurek} proposed a generalization 
of Eq. \eqref{MIC} assuming a complete unidimensional projective measurement 
made on subsystem $B$, with projectors $\{\Pi_j^B\}$ satisfying 
$\sum (\Pi_j^B)^{\dagger}\Pi_j^B=\text{\bf 1}$. The generalization of Eq. \eqref{MIC}  
allows another way of writing the quantum mutual information (which in general does not coincide with that of Eq. \eqref{MI}):
\begin{equation}
   J(\rho_{AB})_{\{\Pi_j^B\}}=S(\rho_A)-S(\rho_{A|\{\Pi_j^B\}}) ,
   \label{QMI}
\end{equation}
where the conditional entropy 
$S(\rho_{A|B})\equiv S(\rho_{A|\{\Pi_j^B\}})=\sum_jp_jS(\rho_{A|\Pi_j^B})$, with probability 
$p_j=tr(\Pi_j^B\rho_{AB}\Pi_j^B)$, and the density matrix after the measurements 
have been performed on subsystem $B$ is given by
\begin{equation}
  \rho_{A|\Pi_j^B}=\frac{\Pi_j^B\rho_{AB}\Pi_j^B}{tr(\Pi_j^B\rho_{AB}\Pi_j^B)} .
   \label{DMAM}
\end{equation}
Equation \eqref{QMI} gives the amount of information 
gained about $A$ after a  measure has been carried out on $B$.

\subsection{Classical correlations} 
\label{secCC}

A measure for classical correlations (CCs) was introduced in Ref. \cite{Vedral}. 
This quantity is defined as the maximum extractable classical information from a subsystem, say  $A$, when a set of positive operator valued measures (POVMs) \cite{man}
has been performed on the other subsystem ($B$):
\begin{eqnarray}
\label{CC}
  CC(\rho_{AB}) &=&\text{max}_{\{\Pi_j^B\}}J(\rho_{AB})_{\{\Pi_j^B\}}  =\\ &&  \nonumber
\text{max}_{\{\Pi_j^B\}}\Big(S(\rho_{A})-\sum_jp_jS(\rho_{A}^j)\Big) ,
\end{eqnarray}
where $S(\rho_{A}^j)$ is the entropy associated to the density matrix of subsystem $A$ after the measure.  Briefly, a measure of classical correlations should  incorporate the following:  i) 
$CC(\rho_{AB}) = 0$ if and only if $\rho_{AB}= \rho_A\otimes \rho_B$, and ii)   $CC$ must be  non-increasing, and  invariant under local unitary operations. In addition,  the set of POVMs  that  maximize the classical correlation Eq. \eqref{CC} is 
a complete unidimensional projective measurement $\{\Pi_j^B\}$  \cite{Hamieh}. If $\{\ket{0},\ket{1}\}$ defines the basis states for qubit  $B$,  the projectors can be written as
$\Pi_j^B=\text{\bf 1}\otimes \proj{j}$,  $j=a,b$, 
where 
$\ket{a}= \cos{\theta}\ket{0}+e^{\text{i}\phi}\sin{\theta}\ket{1}$,  
 $\ket{b}= e^{-\text{i}\phi}\sin{\theta}\ket{0}-\cos{\theta}\ket{1}$,
and the optimization is carried out over angles $\theta$ and $\phi$.

\subsection{Quantum correlations} 

\subsubsection{Quantum Discord} 

Following Sec. \ref{secTC}, and given that Eqs. \eqref{MI} and \eqref{QMI} are not always equivalent, a measure of the quantumness of correlations---the so-called quantum discord (QD) is 
defined as the minimum of the difference 
between the Eqs. \eqref{MI} and \eqref{QMI}. The QD then reads
\begin{eqnarray}
   \label{D}
   D(\rho_{AB})&=&\text{min}_{\{\Pi_j^B\}}\left(I(\rho_{AB})-J(\rho_{AB})_{\{\Pi_j^B\}}\right) = \\ \nonumber &&
S(\rho_B)-S(\rho_{AB})+\text{min}_{\{\Pi_j^B\}} \sum_jp_jS(\rho_{A|\Pi_j^B}),
\end{eqnarray}
and the measure of total correlations becomes 
$I(\rho_{AB})=D(\rho_{AB})+CC(\rho_{AB})$, where $D(\rho_{AB})$ is a measure of `purely quantum' correlations, and  $CC(\rho_{AB})$ is the maximum extractable classical information. Briefly, 
some properties can be inferred from Eq. \eqref{D}: i) For pure states, $D=S(\rho_B)$, ii)  $D=0$ if and only if $\rho_{AB}=\sum_j\Pi_j^B\rho_{AB}\Pi_j^B$, iii) $0\leq D\leq S(\rho_B)$, iv) $D$ is an entanglement monotone on pure states.
 Both QD and CC are antisymmetric by definition, i.e., they 
depend on what subsystem the measures are taken on \cite{luo1,luo2,kavan}. Without loss of generality, we take qubit $B$ to be  the measured one.

\subsubsection{Quantum Entanglement}

Entanglement is a special and a very particular kind of correlation  \cite{bennett}, radically different to the quantum discord. It is fundamental in quantum phenomena, and a precious resource for performing quantum computing  \cite{ladd, knill} and quantum communication protocols \cite{man}. 
A bipartite pure state is said to be \emph{entangled} 
if at least two cofficients of its Schmidt decomposition  do not vanish \cite{man}, and the operators $\rho_{A}$ and $\rho_{B}$ have the 
same non-vanishing eigenvalues: they are equal to the square of the Schmidt 
numbers.
A 
state\index{state} acting on Hilbert space
${\cal H}_{AB}$
is \emph{separable} 
if it is of the form $\rho=\sum_{i=1}^k c_i\rho_i^A\otimes \rho_i^B$, 
for some $k$, where $\rho_i^A$ and $\rho_i^B$ are states\index{state} 
on ${\cal H}_A$ and
${\cal H}_B$ respectively.
If $\rho$ is a {\it pure} state, i.e., 
$\rho=\kett{\Psi_{AB}}\braa{\Psi_{AB}}$, then it is
 separable if and only
if $\kett{\Psi_{AB}}=\kett{\Psi_A}\otimes \kett{\Psi_B}$.

The  {\it entanglement of formation} (EoF), arguably the most prominent  measure of entanglement, allows an analytical expression in  the bipartite case. 
The EoF for the density matrix $\rho$ is defined as the average of the entanglement of all the states composing $\rho$, minimizing over the whole composition (over pure state decompositions) \cite{wootter},
\begin{equation}
  EoF(\rho)=\text {min}\sum_{i}p_i E(\psi_i) ,
\end{equation}
where
$
  E(\psi)=-tr[\rho_A\log_2\rho_A]=-tr[\rho_B\log_2\rho_B]
$
is the entropy of each of the subsystems. 
 The EoF can also be written in terms of the quantum mutual information  as 
$
EoF(\rho_{AB})= \min \sum p_i \,\,\half {I}(\rho^i_A:\rho^i_B).
$
Although the minimization is not trivial in general, the EoF has an analytical solution for the two-qubit case~\cite{wootter}.

For {\it pure} states,  the EoF can be written as
\begin{equation}
  EoF(\psi)={\mathcal E}(C(\psi)) ,
\end{equation}
where the concurrence $C(\psi)=\mid\braket{\psi}{\psi}\mid$, and the function 
\begin{equation}
  {\mathcal E}(C)=h\left(\frac{1+\sqrt{1-C^2}}{2}\right) ,
\label{formation}
\end{equation}
where 
$h(x)=-x\log_2x-(1-x)\log_2(1-x)$ denotes the binary entropy function.  $ {\mathcal E}$ is monotonically increasing, and  it goes from $0$ to $1$ 
when the concurrence $C$ ranges from $0$ to $1$. This is why $C$ can be used as an 
entanglement quantifier. This said, it should be stressed that it is only the EoF 
that is an entanglement measure, and $C$
gets  its meaning via its relation to the EoF.

For {\it mixed} states, the EoF is written as \cite{wootter}
\begin{equation}
  EoF(\rho)={\mathcal E}(C(\rho)) .
  \label{EoFR}
\end{equation}
Consider,  in
decreasing order, the eigenvalues $\lambda_i$ of the matrix
$\sqrt{\rho_{AB} \tilde\rho_{AB}}$, where $\tilde\rho_{AB} =
(\sigma_y\otimes\sigma_y) \bar\rho_{AB} (\sigma_y\otimes\sigma_y)$,  $\bar\rho_{AB}$
is the elementwise complex conjugate of $\rho$, and $\sigma_y$ is the Pauli matrix. 
The concurrence $C$ is defined as 
\begin{equation}
C(\rho_{AB})={\text{max}}\{0, \lambda_1- \lambda_2- \lambda_3-\lambda_4\} , 
\label{concurrence}
\end{equation}
where the $\lambda_i$'s are  as defined above or, equivalently (also in
decreasing order), the square root of the eigenvalues of the non-Hermitian matrix  $\rho_{AB} \tilde\rho_{AB}$  \cite{wootter}. 

\section{Single quantum emitters}
\label{secsystem}

We consider  dipole-dipole (d-d)  interacting quantum 
emitters in contact with a surrounding medium that acts as an environment. This is the case,  for example,  
of coupled terrylene single molecules embedded in a para-terphenyl crystal for which symmetric and anti-symmetric  (super- and sub-radiant) states, as well as d-d interaction strength, and  single and collective decay rates have been measured \cite{Hettich}. For these, molecular quantum control and entangling quantum logic gates have been designed  \cite{jh1}, in addition to the control of the quantum dissipative  entanglement dynamics \cite{jh3, tannnas}. Moreover, quantum control experiments and  femtosecond single-qubit gates involving single organic molecules at room temperature have recently been performed \cite{niek1}.

For the purpose of the analysis of the dynamics of the correlations  presented here,  we consider that the 
 emitters couple with  interaction strength $V$,  and have individual  transition frequencies $\nu_{i}$ (molecule $i$) in 
 interaction with the quantized radiation
field. We denote  by 
$\ket{0_{i}}$, and $\ket{1_{i}}$, $i=1,2$, the ground and excited state of emitter  $i$, respectively, 
and hence the single particle Hamiltonian $\hat{H}_0=-\frac{h}{2}\nu_1\sigma^{(1)}_{z}-\frac{h}{2}\nu_2\sigma^{(2)}_{z}$.
The  two
two-level emitters are separated by the
vector $\mathbf{r}_{12}$ and are characterized by  transition dipole moments
 $ \hat{\boldsymbol{\mu}}_{i}\equiv\bra{0_i}\mathbf{D}_{i}\ket{1_i}$, with dipole operators 
 $\mathbf{D}_{i}$, and spontaneous emission rates $\Gamma_i$.
 The emitters are externally controlled by a coherent driving laser field 
which  acts on each of the molecules with a coupling 
amplitude $h\ell_{i}=-\boldsymbol{\mu}_i\cdot \boldsymbol{E}_i$, and a frequency $\omega_{L}$. 
The light-matter interaction Hamiltonian  $\hat{H}_L=h \ell^{(i)}(\sigma^{(i)}_{-}e^{i\omega_{L}t}+\sigma^{(i)}_{+}e^{-i\omega_{L}t})$, where 
$\boldsymbol{E}_i$ 
is the amplitude of the coherent driving acting on molecule $i$ located at position $\boldsymbol{r}_i$. We also allow for the molecular and laser detunings $\Delta\_\equiv\nu_1-\nu_2$, and $\Delta_{+}\equiv\frac{\nu_{1}+\nu_{2}}{2}-\nu_L$.

\section{Dissipative dynamics of quantum correlations}
\label{secdissipative}
The  emitters interaction Hamiltonian can be written in 
the computational basis of direct product states $\ket{i}\otimes\ket{j}$ $(i, j=0,1)$ as
$\hat{H}_S=\hat{H}_0+\hat{H}_{12}$, where   
$\hat{H}_{12}$ is set by the 
dipole coupled  
molecules of interaction energy  
$  \hat{H}_{12} =\frac{hV}{2}\big(\sigma^{(1)}_{x}\otimes\sigma^{(2)}_{x}+\sigma^{(1)}_{y}\otimes\sigma^{(2)}_{y}$\big).
Hence, the dissipative dynamics corresponding to the ligth-matter interaction \cite{agarwal,tannnas} can be described, with $\hat{H}=\hat{H}_S + \Hat{H}_{L}$, 
by  the quantum master equation
%
\begin{equation}
\label{master}
\hat{\dot\rho}= -\frac{i}{\hbar} \big[\hat{H},\hat{\rho}\big]  + L(\hat{\rho}) ,
\end{equation}
where the dissipative Lindblad 
super-operator reads  \cite{jh3}
\begin{eqnarray}
\nonumber
     L(\hat{\rho}) &=&
      -\frac{\Gamma _{1}}{2}\left( {\hat\rho}
      \sigma^{(1)}_{+}\sigma^{(1)}_{-}+\sigma^{(1)}_{+}\sigma^{(1)}_{-}{\hat\rho}
      -2\sigma^{(1)}_{-}{\hat\rho} \sigma^{(1)}_{+}\right) \\
    &  &
 -\frac{\Gamma _{2}}{2}\left( {\hat\rho}
      \sigma^{(2)}_{+}\sigma^{(2)}_{-}+\sigma^{(2)}_{+}\sigma^{(2)}_{-}{\hat\rho}
      -2\sigma^{(2)}_{-}{\hat\rho} \sigma^{(2)}_{+}\right)  \nonumber \\
    &  &
-\frac{\Gamma _{12}}{2}\left( {\hat\rho}
      \sigma^{(1)}_{+}\sigma^{(2)}_{-}+\sigma^{(1)}_{+}\sigma^{(2)}_{-}{\hat\rho}
      -2\sigma^{(1)}_{-}{\hat\rho} \sigma^{(2)}_{+}\right) \nonumber \\
   &  &
 -\frac{\Gamma _{21}}{2}\left( {\hat\rho}
      \sigma^{(2)}_{+}\sigma^{(1)}_{-}+\sigma^{(2)}_{+}\sigma^{(1)}_{-}{\hat\rho}
      -2\sigma^{(2)}_{-}{\hat\rho} \sigma^{(1)}_{+}\right), \nonumber
\label{diss}
\end{eqnarray} 
where  $\sigma^{(i)}_{+}=\ket{1_{i}}\bra{0_{i}}$, and $\sigma^{(i)}_{-}=\ket{0_{i}}\bra{1_{i}}$ are the raising 
and lowering Pauli operators acting on  emitter $i$.
$\Gamma_{i}$, 
and $\Gamma_{12}=\Gamma^{\ast}_{21}\equiv \gamma$ 
are the individual, 
and the collective spontaneous emission 
rates, respectively. The latter arises  from the coupling between the emitters through the 
vacuum field.
The interaction  strength and the incoherent decay rate read 
\begin{eqnarray}
V&=&
\label{V}
\mathcal{C}
\Big(-[\hat{\boldsymbol{\mu}}_{1}\cdot  \hat{\boldsymbol{\mu}}_{2}
-(\hat{\boldsymbol{\mu}}_{1}\cdot
      \hat{\mathbf{r}}_{12})
(\hat{\boldsymbol{\mu}}_{2}\cdot
      \hat{\mathbf{r}}_{12})] \frac{\cos z}{z} + \\  && \nonumber
 [\hat{\boldsymbol{\mu}}_{1}\cdot  \hat{\boldsymbol{\mu}}_{2}
-3(\hat{\boldsymbol{\mu}}_{1}\cdot
      \hat{\mathbf{r}}_{12})
(\hat{\boldsymbol{\mu}}_{2}\cdot
      \hat{\mathbf{r}}_{12})]
\Big[ \frac{\cos z}{z^{3}}+\frac{\sin z}{z^{2}}\Big]\Big) , \\ 
\label{inco}
\gamma&=&
\mathcal{C}\Big(
[\hat{\boldsymbol{\mu}}_{1}\cdot  \hat{\boldsymbol{\mu}}_{2}
-(\hat{\boldsymbol{\mu}}_{1}\cdot
      \hat{\mathbf{r}}_{12})
(\hat{\boldsymbol{\mu}}_{2}\cdot
      \hat{\mathbf{r}}_{12})] \frac{\sin z}{z} + \\ && \nonumber
[\hat{\boldsymbol{\mu}}_{1}\cdot  \hat{\boldsymbol{\mu}}_{2}
-3(\hat{\boldsymbol{\mu}}_{1}\cdot
      \hat{\mathbf{r}}_{12})
(\hat{\boldsymbol{\mu}}_{2}\cdot
      \hat{\mathbf{r}}_{12})] \Big[\frac{\cos z}{z^{2}}-\frac{\sin z}{z^{3}}\Big]\Big),
\end{eqnarray}
where $z= nk_{0}r_{12}$,  $n$ denotes the medium  refraction index, $k_{0}=\frac{\omega_{0}}{c}$, $\omega_{0}=\frac{\omega_{1}+\omega_{2}}{2}$, and $\mathcal{C}= \frac{3\sqrt{\Gamma_1 \Gamma_2}}{8\pi}$.

\subsection{Identical emitters without laser field}
\label{secidentical}
We begin with the simplest case scenario, that of 
 identical emitters or qubits, $\Delta_{-}=0$, no external laser field:  $\ell_{i}=0$, and 
$\Delta_{+}=(\nu_{1}+\nu_{2})/2$. 

We consider the emitters dipolar moments to be parallel amongst them and perpendicular to the interqubit position vector ${\bf r}_{12}$.  In particular,  if $r_{12}\ll\lambda_0$, the maximum (minimum) interaction strength  and decay rate are obtained for
parallel (perpendicular) dipole moments as 
$
hV = \frac{3h\sqrt{\Gamma_1 \Gamma_2}}{8\pi z^3}\left[\hat{\boldsymbol{\mu}}_{1}\cdot  \hat{\boldsymbol{\mu}}_{2}
-3(\hat{\boldsymbol{\mu}}_{1}\cdot
      \hat{\mathbf{r}}_{12})
(\hat{\boldsymbol{\mu}}_{2}\cdot
      \hat{\mathbf{r}}_{12})\right]$, 
      $\gamma =\sqrt{\Gamma_1 \Gamma_2}\; 
\hat{\boldsymbol{\mu}}_{1}\cdot  \hat{\boldsymbol{\mu}}_{2}$. However, our calculations  make direct use of  Eqs. \eqref{V} and \eqref{inco}, since  we investigate the dependence of correlations on the interqubit separation.

\begin{figure}[htb]
  \centering
  \includegraphics[width=8.0cm]{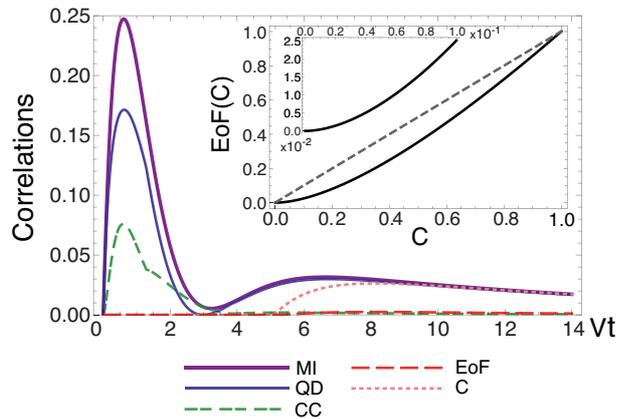}
  \caption{(Color online) Correlations dynamics for a pair of interacting emitters initially in the doubly-excited state $\ket{11}$.
  MI (solid thick line, purple), QD (solid thin line, blue), CC (doubly  dashed line, green),  
  C (dotted  line, pink), EoF (dashed  line, red). $\gamma=0.6884 \, V$. Main: $\Gamma=0.7818 \, V$ ($r_{12}=\lambda_0/8$). Inset: Functional relation between EoF and C. These functions are monotonically 
  increasing from $0$ to $1$. The  zoom points out the differences between the EoF ($\in[ 0,0.025]$) and  C ($\in[ 0,0.1]$). }
  \label{fig1}
\end{figure}

The effect of dissipation on the correlations dynamics is initially shown in Fig. \ref{fig1} for the decay rate $\gamma=0.6884 \, V$, and  $\Gamma_1=\Gamma_2\equiv\Gamma$. Here, weakly coupled qubits are considered for interaction strengths of the order of the decay rates $V\sim\Gamma$ which in the main graph corresponds to $r_{12}=\lambda_0/8$.  The correlations last longer times and reach higher values
the smaller $\Gamma$.  As expected, the quantum mutual information is always greater than the rest of the correlations. For the emitters in the initial doubly excited state $\ket{11}$ we find that the entanglement quantifiers EoF and C exhibit  a `sudden birth' of entanglement \cite{jh3,eberly} at a time $\tau\sim 5/V$ (main graph); this behavior is `shortened' as the decay rate increases (not shown). In contrast to this, the quantum discord is always greater than entanglement (measured by EoF and C), and only vanishes around a time $3/V$: only around a small time neighbourhood centered at this value do the classical correlations (CC) become slightly greater than quantum correlations (QD).  For the remaining parts of the evolution,  QD $>$ CC. We shall return to this point later, where we address the question of whether it is at all feasible to characterize a dynamical order relation between classical and quantum correlations. We remark  that even in the absence of entanglement ($t<\tau$) there are both, quantum and classical correlations present in the dynamics, and that the QD is the only quantum correlation present in such a time frame. 

\subsection{Entanglement of Formation vs Concurrence}

The concurrence has been widely used as an entanglement metric due to the fact that,  according to Eq. \eqref{EoFR}, the function EoF(C) is monotonically increasing and ranges from $0$ to 
$1$ as C goes from $0$ to $1$.
This said, the difference in value between these 
quantifiers can be significant and this is reflected in the interpretation of the correlation dynamics. 
The 
EoF changes at a slower rate than C  at values close to $0$; for instance, in the interval $C\in[ 0,0.1]$,  the EoF is much smaller 
than $C$, as can be seen in the zoom of the inset of Fig. \ref{fig1}. The whole range for the metric  is plotted in the inset of Fig.  \ref{fig1}, where the dashed line is a guide to the eye and indicates the equality EoF $=$ C, to  better appreciate the differences between EoF and C (solid curve).

As for the physical interpretation of the correlations dynamics, we relate the above discussion with that  presented in Fig. \ref{fig1}. 
Here, the dotted (pink) curve denoting C gives the impression that entanglement is almost the only quantum correlation present in the system for times around or greater than twice the sudden birth time  ($t>2\tau$), since for this, C and QD are almost, or actually, superimposed.  The solid (red) curve for 
the EoF clearly shows that most quantum 
correlations, however, in the system are far from entanglement. As a matter of fact, entanglement, as measured by EoF, contributes very little to the full quantum correlations dynamics.  
\begin{figure}[t]
  \centering
  \includegraphics[width=8cm]{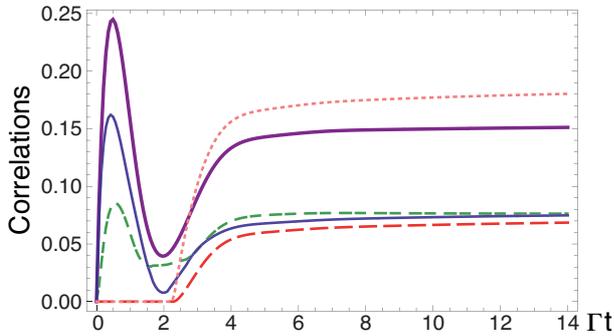}
  \caption{(Color online) Same parameters and notations  as in Fig. \ref{fig1}  (main), but in the presence of an external continuous laser field of amplitude $l=0.4\Gamma$ in  resonance with the qubits transition energies: $\Delta_+=0$ ($\omega_L=\omega_1=\omega_2\equiv\omega$). The initial condition is  taken to be the doubly-excited state $\ket{11}$.}
  \label{fig1l}
\end{figure}

This point is made even more evident when we consider, for example, the influence of an external continuous wave laser field of amplitude $l$, and  frequency $\omega_L$ in resonance with the qubits transition energies: $\omega_L=\omega$ (below we shall return to the discussion of the laser effects),  as shown in Fig. \ref{fig1l}. The physical parameters and the notations are the same as those shown in the main graph in Fig. \ref{fig1}. The result that marks the importance of using EoF as the metric of entanglement is that, while in Fig. \ref{fig1l} the EoF (red curve, just  below the discord which is plotted in blue) shows a behavior that can be understood following the lines of reasoning in  Fig. \ref{fig1} (but now in the presence of a laser field; note the differences), the concurrence (dotted pink curve) exhibits a `birth' of entanglement (following the qualitative behavior of the EoF), just around a time that marks the beginning of the dominance of classical over quantum correlations,  that grows well beyond the mutual information: since the latter is, by definition, the measure of total correlations present in the system at any given time,  concurrence cannot be a good quantifier of entanglement in  this context.
This is an important point to bear in mind when performing quantum information protocols, where entanglement is required as a physical resource. 
As already pointed out, it is only the EoF 
that is an entanglement measure, and the significance of C
is actually revealed through its relation to the EoF. Hereafter, we shall consider the EoF as
the quantifier of the emitters' entanglement.

\section{Hierarchy for the emitters classical and quantum correlations}
\label{secorder}

Regarding the conjecture that classical correlations are always greater than quantum correlations, and the recent experimental evidence that the opposite situation can also arise for a specific scenario in bipartite entangled photons subjected to a decoherent source (thus disproving the conjecture) \cite{exp1}, we ask whether it is at all feasible, and under what circumstances, to find and order hierarchy for the emitters' classical and quantum correlations.

The following  analysis of  dynamical behavior for the correlations is given in terms of the inherent physical properties of the emitters, such as   
the collective damping  (decay) rate $\gamma$, and the d-d interaction $V$. We demonstrate, in a simple manner, that the dimer system can be naturally dominated 
by quantum correlations {\it at all times}, that is, that QD or EoF  are greater than classical correlations during the whole time evolution. 
 For this, we solve the emitters' master equation \eqref{master}
for identical qubits ($\nu_1=\nu_2$), with  emitters only coupled through their dipolar interaction (there is no laser excitation).
The calculation gives, for the non-trivial density matrix elements, the following result:
\begin{eqnarray}
\nonumber 
\rho_{00,00} &=& \frac{1}{2} \mathrm{e}^{-(\gamma+\Gamma )t} \Big[2 \mathrm{e}^{ (\gamma +\Gamma)t}-2 f (1-\mathrm{e}^{2  \gamma t})\cos{\varphi}\\ \nonumber
& & -\mathrm{e}^{2 \gamma t}-1\Big] \\ \nonumber 
\rho_{01,01} &=&\frac{1}{4} \mathrm{e}^{- \Gamma t } \Big[\mathrm{e}^{- \gamma t} \big(1+\mathrm{e}^{2\gamma t}+2 f(1-\mathrm{e}^{2  \gamma t})\cos{\varphi}\big)\\ \nonumber
& & - (2-4\alpha) \cos{\Theta}+4 f \sin{\varphi}\sin{\Theta}\Big] \\ \nonumber
\rho_{10,10} & = &\frac{1}{4} \mathrm{e}^{- \Gamma t } \Big[\mathrm{e}^{- \gamma t} \big(1+\mathrm{e}^{2\gamma t}+2 f(1-\mathrm{e}^{2  \gamma t})\cos{\varphi}\big)\\ \nonumber
& & + (2-4\alpha) \cos{\Theta}-4 f \sin{\varphi}\sin{\Theta}\Big] \\ \nonumber
\rho_{01,10}&=&\rho_{10,01}^{\ast} =\\ \nonumber
& & \frac{1}{4} \mathrm{e}^{- \Gamma t } \Big[\mathrm{e}^{- \gamma t} \big(1-\mathrm{e}^{2\gamma t}+2 f(1+\mathrm{e}^{2  \gamma t})\cos{\varphi}\big)\\
& & - 2\mathrm{i}\big(2 f \sin{\varphi}\cos{\Theta}+(1-2\alpha) \sin{\Theta}\big)\Big] \, ,
\label{sims}
\end{eqnarray}
where $f=\sqrt{(1-\alpha) \alpha}$, $\Theta=2Vt$,  $\varphi$ is a relative phase, and $\alpha$ defines the set of 
initial states $\ket{\alpha_{\varphi}}\equiv \sqrt{\alpha}\ket{01}+\mathrm{e}^{\mathrm{i}\varphi}\sqrt{1-\alpha}\ket{10}$,  $0\leq \alpha\leq 1$, 
a choice motivated  from states that can be `naturally' prepared in the quantum  emitters, e.g., separable states $\alpha=0,1$, and 
symmetric ($\alpha=1/2$,  and $\varphi=0$) and anti-symmetric 
($\alpha=1/2$,  and $\varphi=\pi$) entangled states. 
For the class of $\alpha_{\varphi}$ initial states, whose time-evolution has been derived in Eq. \eqref{sims}, 
the  density matrix adopts  the structure
%
\begin{eqnarray}
\rho(t)=\left(
\begin{array}{cccc}
\rho_{00,00}  & 0 & 0 & 0 \\
0 & \rho_{01,01} & \rho_{01,10} & 0 \\
0 & \rho_{01,10}^{\ast} & \rho_{10,10} & 0 \\
0 & 0& 0 & 0
\end{array}\right) ,
\label{rhosim}
\end{eqnarray}
and thus the correlations can be analytically calculated. In particular, we are interested in finding a dynamical  order hierarchy amongst  classical and quantum correlations. Equations \eqref{CC} and \eqref{D} allow to show that there is an entropy bound:
\begin{equation}
D(\rho_{AB})\geq CC(\rho_{AB}),
\end{equation}
which reads
\begin{equation}
  S(\rho_{B})+2 \text{min}_{B_{j}}\{S(\rho_{A|\{\Pi_j^B\}})\}-S(\rho_{A})-S(\rho_{AB})\geq 0 .
  \label{enc}
\end{equation}
When the system begins in the 
maximally entangled symmetric Bell state $\alpha =\frac{1}{2}, \, \varphi=0$, the density matrix becomes {\it independent} of 
the dipolar interaction $V$, and $\gamma$ is just an additional decay rate that makes the system decay with the new 
 rate $\Gamma +\gamma$; furthermore, $S(\rho_{A})=S(\rho_{B})$. 

\begin{figure}[t]
  \centering
  \includegraphics[width=7.5cm]{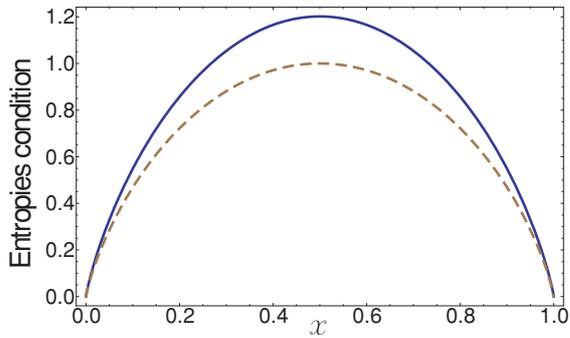}
  \caption{(Color online) Hierarchy of correlations: $D(\rho_{AB})\geq CC(\rho_{AB})$ {\it for all} $t$. $x={\rm exp}[-(\Gamma+\gamma) t]$. The top (solid) curve is the quantity 
  $2 \text{min}_{B_{j}}\{S(\rho_{A|\{\Pi_j^B\}})\}$, and the bottom (dashed) curve corresponds to the entropy $S(\rho_{AB})$. $\alpha=1/2$, $\varphi=0$; $S(\rho_{A})=S(\rho_{B})$.}
  \label{figQC1}
\end{figure}

The independence of $\hat\rho$ with respect to $V$ can be formally understood by recalling that  the considered Bell states are part of a 
bigger family of states---$\rho_M$, which have   maximally mixed marginals, i.e.,  $tr_A(\rho_{M})=tr_B(\rho_{M})=\frac{1}{2}I_{2\times2}$,  
with $I_{2\times2}$ being the identity $2\times2$ matrix; hence $S(\rho_{A})=S(\rho_{B})=1$. Such  states can be written as 
$
\rho_{M}=\frac{1}{4}\big(I+\sum_{i=1}^3h_i\sigma_i\otimes\sigma_i\big) 
$
 \cite{luo2,fano}, where $\sigma_i$ are the Pauli matrices and the  coefficients $h_i\in \cal{R}$ satisfy constraints such that $\rho_M$ 
is a well defined density operator. It can be shown that for the whole family of density matrices $\rho_M$, the Markovian time 
evolution given by Eq. \eqref{master} is completely independent of the interaction $V$.

Returning to our case,  quantum correlations are {\it always greater}   than  classical correlations. This can be seen in 
Fig. \ref{figQC1}, where we have plotted the entropies condition that make $D(\rho_{AB})\geq CC(\rho_{AB})$. With the parametrization  $x=\exp[{-(\Gamma+\gamma) t}]$, in  Fig. \ref{figQC1} we show that the quantum discord is greater 
than classical correlations for the whole dynamics. The equality is only obtained at the initial time 
(pure state) and at
the `final'  ($t\rightarrow\infty$) time; in the latter, the system tends to a separable and uncorrelated state. In a similar manner, we can also show that the entanglement of formation $EoF(\rho_{AB})$ is always greater than the classical correlations.

The same hierarchy of correlations can be proven by direct calculation. For the density matrix \eqref{rhosim}, the 
conditional entropy, after the measurement, admits an analytical expression and therefore  both 
$CC$ and $D$ can be analytically computed:  the conditional entropy has  two possible 
solutions:
\begin{eqnarray}
\nonumber 
S^{(1)}(\rho_{A|\{\Pi_j^B\}}) &=&-\rho_{00,00} {\text {log}}_2 \left(\frac{\rho_{00,00}}{\rho_{00,00}+\rho_{10,10}}\right) \\ \nonumber
& &   -\rho_{10,10} {\text {log}}_2\left(\frac{\rho_{10,10}}{\rho_{00,00}+\rho_{10,10}}\right), \\ \nonumber
& & \\ \nonumber 
S^{(2)}(\rho_{A|\{\Pi_j^B\}})&=&-\frac{1}{2}(1-\xi) {\text {log}}_2\left(\frac{1}{2}(1-\xi)\right)\\ 
& &  -\frac{1}{2}(1+\xi) {\text {log}}_2\left(\frac{1}{2}(1+\xi)\right),
\label{twos}
\end{eqnarray}
where $\xi=\sqrt{(1-2\rho_{10,10})^2+4|\rho_{01,10}|^2}$. Then, $D$ and $CC$ are 
determined by the minimum of the two expressions in (\ref{twos}). 

For  $\alpha=\frac{1}{2}$, we find  that  $S^{(2)}(\rho_{A|\{\Pi_j^B\}})\leq S^{(1)}(\rho_{A|\{\Pi_j^B\}})$ is always satisfied and thus $S^{(2)}(\rho_{A|\{\Pi_j^B\}})$ is the entropy used for computing the correlations. 

The  concurrence for  the 
density matrix (\ref{rhosim}) takes the simple form $C(\rho)=2\, {\rm max}\{0, \left|\rho_{01,10}\right|\}$. The entanglement of formation is then  directly computed from Eq.~\eqref{EoFR}.
\begin{figure}[htb]
  \centering
  \includegraphics[width=7cm]{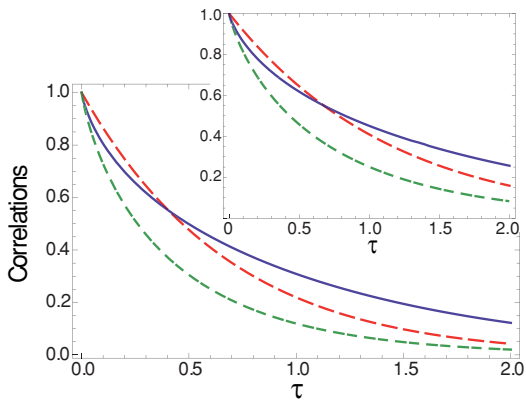}
  \caption{(Color online) Dynamics of correlations. $D(\rho_{AB})$ is the solid curve, the  $EoF(\rho_{AB})$ is the dashed curve, and $CC(\rho_{AB})$  the doubly dashed curve.  The initial state is given by the super-radiant state 
  $\alpha_{\varphi=0}=1/2$ (main), and the sub-radiant state  $\alpha_{\varphi=\pi}=1/2$  (inset).}
  \label{figQC2}
\end{figure}

In Figure \ref{figQC2} we show the evolution of these three correlations as a function of the normalized time 
$\tau=(\Gamma+\gamma) t$. The exponential decay is due to the particular choice of maximally entangled initial states  
  $\alpha=1/2$ for (a) the symmetric state $\varphi=0$ (main), and (b) the anti-symmetric state $\varphi=\pi$ (inset). Although the  qualitative behavior of the inset is similar to that of the main graph, we notice  that such state evolution has a much slower  decay rate of the correlations; this happens at the new rate $\Gamma-\gamma$. This result is in agreement with what would be expected for sub-radiant states and explains the higher value of the correlations at all times, when compared to the  initial super-radiant  
$\varphi=0$ state.
\begin{figure}[htb]
  \centering
  \includegraphics[width=7.5cm]{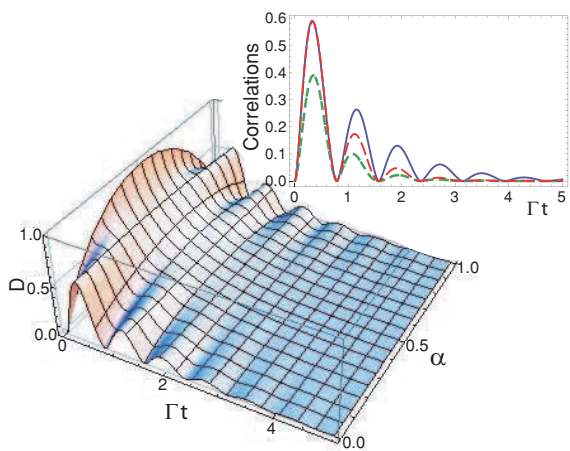}
  \caption{(Color online) Quantum discord dynamics as a function of the initial   $\alpha_{\varphi=0}$ state preparations.   The inset shows the dynamics of correlations $D(\rho_{AB})$ (solid blue),  $EoF(\rho_{AB})$ (dashed, red), and $CC(\rho_{AB})$ (doubly dashed, green),  for the (separable state) $\alpha=0$-plane. There is no collective damping: $\gamma=0$, and  $V=2 \, \Gamma$.}
  \label{figQC3}
\end{figure}

\begin{figure}[htb]
  \centering
  \includegraphics[width=7.5cm]{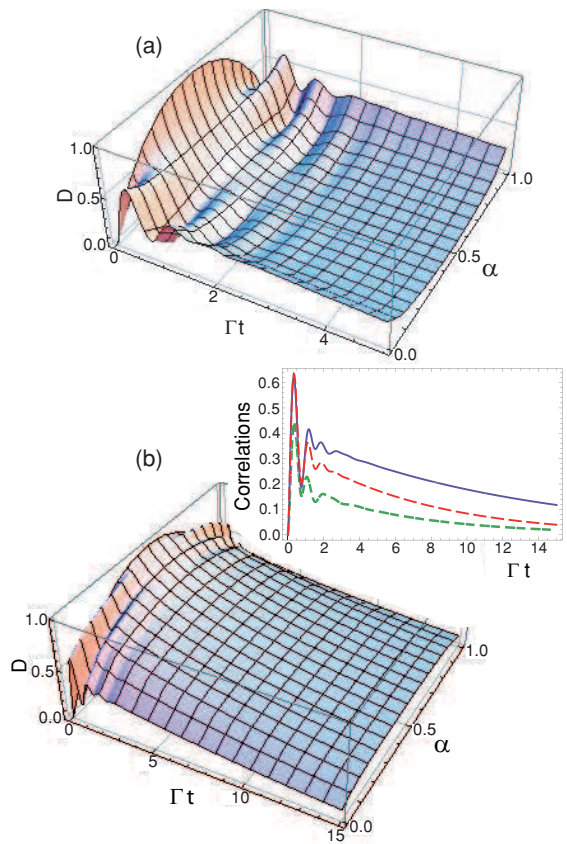}
  \caption{(Color online) 
Quantum discord dynamics as a function of the initial state preparations (a)  $\alpha_{\varphi=0}$, and (b)   $\alpha_{\varphi=\pi}$ states (main).  $\alpha=1/2$ give the super-radiant and sub-radiant states, respectively. The inset notations are as in Fig. \ref{figQC3}, for $\alpha=0$. 
 $\gamma=0.91\,\Gamma$, and $V=2.03 \,\Gamma$.}
  \label{figQC4}
\end{figure}

 The above analysis is extended to all initial states $0\leq\alpha\leq 1$, see Figs. \ref{figQC3} and \ref{figQC4}. We also find, for  identical qubits, that quantum correlations are stronger and more 
robust to decay than  classical correlations. The dynamical behavior is richer than the exponential decay of Fig. \ref{figQC2} because different initial conditions for the density matrix allow for different roles for the corresponding collective parameters.
For $\alpha\neq 1/2$, the last terms of matrix elements $\rho_{01,01}$ 
and $\rho_{10,10}$, and the imaginary part of $\rho_{01,10}$, are non-zero and hence the dynamics exhibits oscillations depending on the value of the d-d interaction,  which  is weak for the case of the diluted molecules first considered; accordingly,  we consider in the  first place  the scenario of no collective decay due to $\gamma$ (decay contributions are only those giving by $\Gamma$), as shown in Fig. \ref{figQC3} for the initial state dynamics
  $\alpha_{\varphi=0}$.  We also consider, in Fig. \ref{figQC4}, the situation where $\gamma\neq 0$ plays an important role, for the initial state preparations (a) $\alpha_{\varphi=0}$, and (b)   $\alpha_{\varphi=\pi}$. The correlation dynamics for these two cases is shown in Figs. \ref{figQC3} and \ref{figQC4}.

When the collective decay rate is not effectively present in the dynamics,  as, for example, in the case of diluted molecules in a dispersive environment for which their collective rate is much less than both the spontaneous individual emission rates and the dipolar interaction \cite{Hettich,niek1}; or in  a pair of 
emitters that interact with a plasmonic channel where  the collective stregths $V$ and $\gamma$ can effectively be switched on and off 
as the distance between the emitters is varied, a mechanism that produces an effective  phase difference between $V$ and $\gamma$ \cite{tudela}.

The main graph in Fig. \ref{figQC3} plots the quantum discord dynamics as a function of $\alpha$ for two different  set of initial state preparations; the same qualitative  behavior is observed for CC and EoF (not shown). 
In Fig. \ref{figQC3} we see that oscillations due to the d-d interaction  begin to appear and to coherently 
 affect the dynamics of the quantum discord for $\alpha$ values closer to $0$ or $1$; this is so   because in these cases 
the amplitude of the trigonometric functions in the matrix elements Eq. \eqref{sims} are closer to their maxima. The inset in Fig. \ref{figQC3} shows this for the discord, the EoF and CCs for the plane $\alpha=0$ (separable state $\ket{10}$). For the maximally entangled state ($\alpha=1/2$) all the correlations have 
a monotonic exponential decay, indicating an independence from the interaction $V$. For other  initial states set by $\alpha$, this decay exhibits oscillations which 
increase their amplitude as  $\alpha$ tends to the separable states $\ket{01}$ or $\ket{10}$. The oscillations eventually vanish and the correlations decay to $0$ for 
all  $\alpha$. Although  the main graphs of  Fig. \ref{figQC3} only plot the discord, our calculations show that the correlations hierarchy  $CC<EoF<QD$ holds for all time $t$ and for all  $\alpha$ (see the inset for the case $\alpha=0$).

The case  $\gamma\neq 0$ (Fig. \ref{figQC4}) plays an important role in the system's evolution. Here,  two 
emitters interact via a dipole force throughout all modes of the electromagnetic field in the vacuum 
state: the collective (incoherent) decay $\gamma$ \eqref{inco} has to be taken into account, and we introduce the full expressions given for $\gamma$ and $V$, and their relation to the
Einstein coefficients, dipole orientation, and the dispersive medium, as given by  Eqs.  \eqref{V} and \eqref{inco}. When compared to Fig. \ref{figQC3},  the effect of the collective decay rate translates into a much longer time for the survival of correlations. In fact, the inset, which is common to Figs. \ref{figQC4}(a) and (b),  shows that quantum correlations decay very slowly and thus last for  a  much longer time  than that obtained in the absence of collective decay in Fig. \ref{figQC3}.
This said,  in Fig. \ref{figQC4}
there is a clear distinction between the initial  (a) $\alpha_{\varphi=0}$, and (b)   $\alpha_{\varphi=\pi}$ states. In the latter, the correlations last for a far longer time before their final decay;  in particular,  the case of the anti-symmetric $\alpha_{\varphi=\pi}=1/2$ state  exhibits a very slow decay of the discord, as compared to the case of  the symmetric $\alpha_{\varphi=0}=1/2$ state (note the difference in the time scale of the two graphs). This is also reflected in the inset of Fig. \ref{figQC4}(b), for the case $\alpha=0$. The difference in physical behavior between graphs (a) and (b) owes its  origin to the nature of the  super-radiance phenomenon, which is manifest here in graph (b) for $\alpha=1/2$. In any case, in both graphs, we have quantum correlations always greater than classical correlations.
\begin{figure}[htb]
  \centering
  \includegraphics[width=6.5cm]{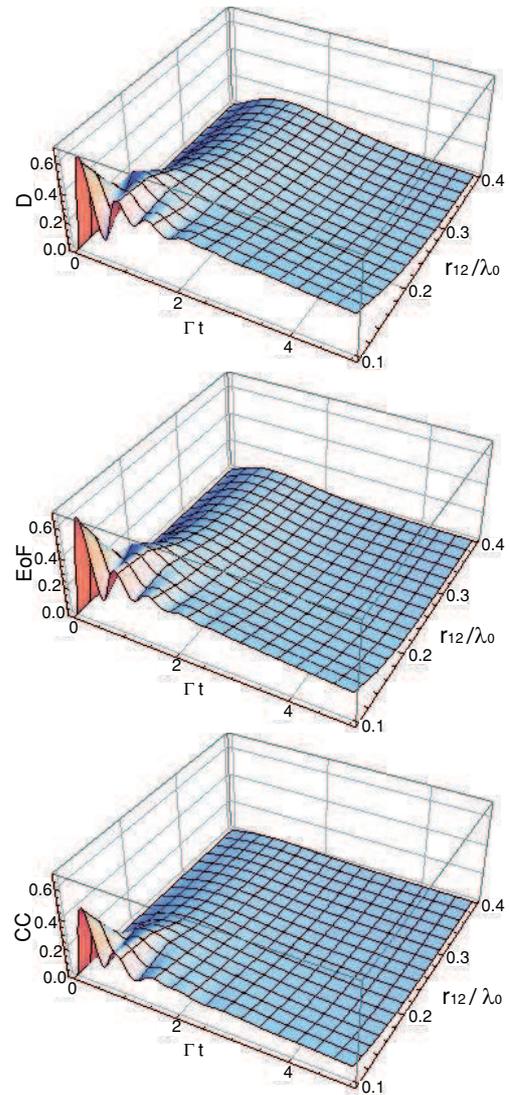}
  \caption{(Color online) Correlations $D$,  $EoF$, and $CC$, as a function of time and the interqubit distance, the latter normalised by the emitters characteristic wavelength $\lambda_0=2\pi/k_0=2\pi c/\omega_0$.  The initial state preparation  corresponds to $\alpha=0$.}
  \label{figQC5}
\end{figure}

 By computing the correlation dynamics for two emitters of identical dipole moments (parallel to each other), and perpendicular to the interqubit  distance ${\bf r}_{12}$,   basic features can be identified. The relevant values of the interaction energy and collective decay rate  are given by  Eqs. \eqref{V} and \eqref{inco}: in Fig. \ref{figQC4} we consider  $\gamma=0.91\, \Gamma$ and $V=2.03\, \Gamma$, which 
corresponds to a distance $r_{12}=0.108\,\lambda_0$. Even though we now  have less  oscillations in the discord,  a higher degree of correlations than those of Fig. \ref{figQC3} is obtained due to the presence of 
collective damping. Note that for initial configurations of the type $\alpha\rightarrow1$ or $\alpha\rightarrow0$, quantum discord and entanglement hold a greater value than  classical correlations and, as can be seen from Figs. \ref{figQC3} and \ref{figQC4}, the  discord is the correlation that most  benefits  due to this collective effect. The {\it entropy condition} Eq. \eqref{enc} {\it  is always} satisfied, and hence  quantum correlations continue to be always 
greater  than classical correlations: $CC<EoF<QD$,  for all  $\alpha$.

According to the above inspection,  a calculation about the way  the collective parameters influence the dynamical behavior 
of correlations is in order. From Eqs. (\ref{sims}) it is clear that a variation of $V$ produces a change in the 
oscillations of the density matrix elements and therefore in the correlations dynamics:  the degree of correlations is increased 
as the collective damping is augmented. Figure \ref{figQC5} shows the behavior of the three relevant correlations as 
a function of the normalized distance $r_{12}/\lambda_{0}$, in the  regime where $V$ goes from $0.2\Gamma$ to $2.6\Gamma$, and $0.1\leq\frac{\gamma}{\Gamma}\leq0.9$, or, equivalently,  for  $0.1\leq\frac{r_{12}}{\lambda_{0}}\leq0.4$. In Fig. \ref{figQC5} we 
consider the initial condition $\alpha=0$,  following the  results shown in Figs. \ref{figQC3} and \ref{figQC4}, as  such a state makes the strongest case for the influence of the collective parameters. As expected, a signature  in the correlations dynamics shows up after the interaction strength becomes either of the order of or greater 
than the decay rate $\Gamma$, which occurs for $\frac{r_{12}}{\lambda_{0}}\leq 0.2$. As in the previous case, the most robust correlation is the quantum 
discord and the classical correlation is the least assisted by the collective effects throughout the entire dynamics. Hence, the degree of 
correlations can be dynamically enhanced when the collective (incoherent) damping 
is of the  order of the (weak) interaction energy and comparable to the spontaneous decay rates of the single quantum emitters.

\begin{figure}[]
  \centering
  \includegraphics[width=6.8cm]{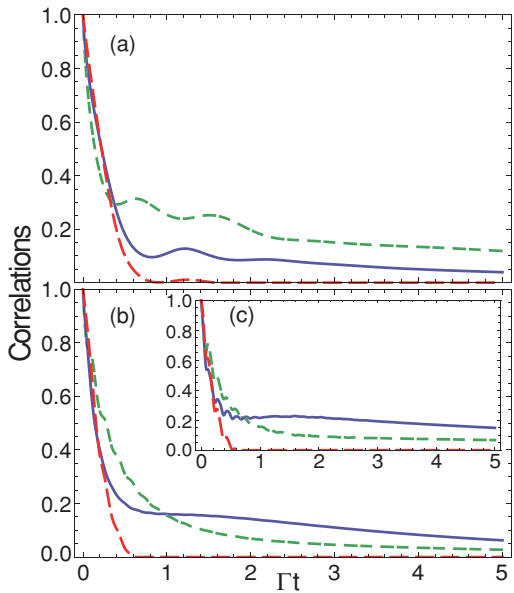}
  \caption{(Color online) Dynamics of correlations for emitters under resonant  laser field excitation: $\nu_{L}=\nu_{0}=\nu$; $\ell_{i}=\ell$. Notations for $D$,  $EoF$, and $CC$  are as in previous figures.
  (a) $\ell=1.5\, \Gamma$, $\gamma=0.91\, \Gamma$, $V=2.03\, \Gamma$; (b) $\ell=10.0\, \Gamma$, $\gamma=0.91\, \Gamma$, $V=2.03\, \Gamma$; (c) $\ell=10.0\, \, \Gamma$, $\gamma=0.97\, \Gamma$, $V=10.45\, \Gamma$. The initial  state is the  super-radiant 
$\alpha_{\varphi=0}=1/2$. }
  \label{figQC6}
\end{figure}
Lindblad \cite{lindblad} conjectured that for a general density matrix $\rho$, classical correlations are  greater than half of the 
mutual information and as such, they are always greater than quantum correlations. He  showed that for an incoherent mixture 
state, $CC(\rho)=I(\rho)$, and that for a pure state,  $CC(\rho)=\frac{1}{2}I(\rho)$,  it seemed  natural
to conjecture that $CC(\rho)\geq \frac{1}{2}I(\rho)$ \cite{lindblad}. Simple numerical examples,  for several  different configurations of the constants $h_i$ in $\rho_M$, confirm that for the cases of incoherent mixtures and pure states, this result holds. There are states for which the conjecture does not hold, however: it is straightforward to show that for the set of Bell states  $|h_i|=1$ (pure correlated states), $I(\rho_{Bell})=2$, hence $CC(\rho_{Bell})=1$. On the other hand, for $h_1=h_2=0$, and $|h_3|\leq1$ (incoherent mixture),  $CC(\rho_{incoh})=I(\rho_{incoh})$. For example, for  $h_3=0.6$, $CC(\rho_{incoh})=I(\rho_{incoh})=0.278$. Finally, the choice  $h_1=h_2=0.8$, $h_3=-0.6$ gives $I(\rho_M)=1.078$, and quantum correlations  $D(\rho_M)=0.547> CC(\rho_M)=0.531$.

\begin{figure}[h]
  \centering
  \includegraphics[width=7.0cm]{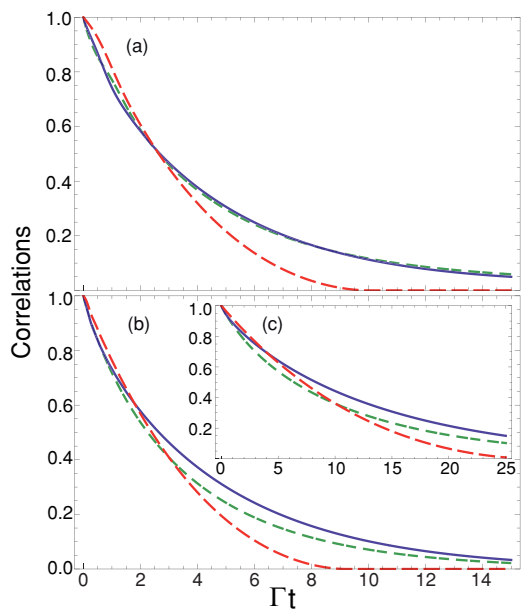}
  \caption{(Color online) Emitters correlations (notation as in Fig. \ref{figQC6}) under resonant laser excitation $\nu_{L}=\nu$.  Parameters are as in Fig. \ref{figQC6}.  Initial  state:  antisymmetric sub-radiant $\alpha_{\varphi=\pi}=1/2$.}
  \label{figQC6a}
\end{figure}

\section{Emitters  correlations under coherent laser excitation} 
\label{seclaser}

We now turn to the external all-optical control of the  quantum emitters   correlation dynamics. We consider  the case of a  continuous laser field 
$\ell_{i} \cos{\omega_{L} t}$ locally  applied to each of the emitters, where  $\ell_{i}$ denotes the amplitude of the 
laser field acting on qubit $i$, and $\omega_{L}$ is the laser frequency. Without loss of generality, we consider  qubits of the same transition frequency $\omega$,  under the resonant condition $\omega_{L}=\omega_{0}=\omega$.

The solution of the master equation describing 
the dynamical system is fully numerical  because the  structure of the density matrix, when  taking into account  the laser field, does not admit a general  analytical solution.

Figure \ref{figQC6} shows the results for initial  state preparation in the super-radiant 
$\frac{1}{\sqrt{2}}(\ket{01}+\ket{10})$ state.
For  external laser  field amplitude  below the interaction energy $V$ (Fig.  \ref{figQC6}(a), and $\gamma$, and $V$ as in Fig. \ref{figQC4}), the emitters dynamics exhibits a change in the correlations hierarchy and, after a brief transient period, quantum  correlations go below the classical ones. This change in physical behavior,  for weak laser excitation, occurs in almost the whole dynamics.
Even though such  laser excitation can overdamp the evolution  of the quantum correlations well below the classical correlations, such evolution can 
be controlled and the degree of the quantum correlations can be optimised  by appropriately tailoring both the amplitude of the field and the dipolar interaction 
between the qubits;  the latter being obtained by varying  the interqubit distance. 
The results plotted  in Fig.  \ref{figQC6a}(a),   for the  initial  state preparation in  the sub-radiant 
$\frac{1}{\sqrt{2}}(\ket{01}-\ket{10})$ state  are in clear contrast with those of 
Fig. \ref{figQC6}(a), and, although quantum correlations are no longer greater than classical correlations, they become of the same order throughout the entire dynamics. The early stage disentanglement (ESD) experienced by the emitters in Fig. \ref{figQC6a}(a) is delayed for about ten times that of Fig. \ref{figQC6}(a). It is worth pointing out that even though there is no entanglement after ESD occurs, from this point on there are non-zero classical and  quantum correlations present in the evolution. The order hierarchy for the anti-symmetric state is such that $QD>CC$ at all times. This is not always the case for the symmetric state, and the hierarchy depends on  the interplay between $V$ and $\ell$.

In Figs.   \ref{figQC6}(b) and  \ref{figQC6a}(b) 
we have kept the interaction energy constant  (fixed collective parameters $V$, and $\gamma$) and increased the value for the 
laser amplitude well above $V$ ($\ell=10\Gamma$). Now  the discord  evolves as  the most robust correlation.  For the case of the symmetric state, however, not always $QD>CC$ (initially, $CC>QD$), whereas for the anti-symmetric state, we always have $QD>CC$. The dynamics becomes more complex with the inclusion of the laser field and a  dependence with the d-d 
interaction now appears,  in contrast to the  case of no laser excitation.  This is seen in Figs. \ref{figQC6}(b) and (c):  for a fixed laser amplitude, the variation in the d-d interaction produces a qualitative change which favours the degree of discord,  above classical correlations and entanglement,  as $V$ increases. The early stage disentanglement doesn't show major changes. 
Now we turn to Figs. \ref{figQC6a}(b) and (c): quantum correlations are always higher than classical correlations and the effect of an increment in $V$ noticeably  prolongues the lifetime and the degree of correlations. For example, ESD in graph \ref{figQC6a}(c) is around 50 times that of \ref{figQC6}(c) (notice the difference in the corresponding  time scales).

In short, the same laser field can reverse the behavior observed in Figs. \ref{figQC6}(a) and \ref{figQC6a}(a) by appropriately changing the interqubit distance and by shining the laser with higher intensity. This control procedure  replicates  the earlier result of quantum correlations being above classical correlations.
 It is also clear that at all times, quantum correlations are far greater than entanglement. The  presence of the external field perturbs the asymptotic decay of entanglement and in turn produces an early stage disentanglement 
 phenomenon \cite{jh3,eberly} which is verified in all the graphs of Figs. \ref{figQC6} and \ref{figQC6a}. 
\begin{figure}[htb]
  \centering
  \includegraphics[width=7.5cm]{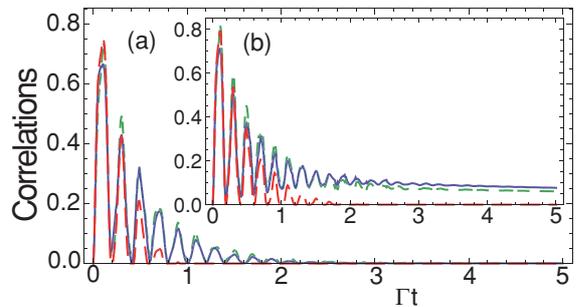}
  \caption{(Color online) Correlations dynamics under the effect of an on-resonance external field $\omega_{L}=\omega_{0}$, 
 with  amplitude $\ell=10\, \Gamma$; $V=10.45\, \Gamma$. (a) Emitters with collective damping `switched off'; 
(b) Emitters with identical and parallel dipoles, with  $\gamma=0.97\, \Gamma$. Initial state: $\alpha=0$.}
  \label{figQC7}
\end{figure}

For completeness, in Fig. \ref{figQC7} we compare the graphs for a separable state as the initial condition, and  for  (a) `absent' $\gamma$ and (b) `active' $\gamma$.   The correlations dynamics has been plotted for $V=10.45\, \Gamma$, with laser amplitude $\ell \sim V$,  following the parameters of graph  \ref{figQC6}(c) which give a high degree of discord for  symmetric  states. Unlike in the previous cases, we now obtain oscillations in the discord and entanglement. Stationary correlations, which are null in the absence of collective damping  (see Fig. \ref{figQC7}(a)), can now be obtained due to the presence of damping $\gamma$ in Fig. \ref{figQC7}(b). Interestingly,  as the correlations stabilise, a quantum discord more robust  than classical correlations emerges.  
This kind  of quantum information or correlations is of an entirely different nature to entanglement, which decreases to zero much earlier than QD and CC.
We  conclude that also for quantum emitters under laser pumping, quantum correlations can become  greater than classical correlations and that this feature is more `natural' in antisymmetric sub-radiant states, thus disproving Linblad's conjecture \cite{lindblad}.

\section{Summary}

We have given a thorough analysis of the dissipative  dynamics of total, quantum, and classical correlations for a set of interacting quantum emitters characterized by their coupling energy and their collective and individual decay rates, and analytically demonstrated the conditions under which quantum correlations are {\it always} greater than classical correlations, and the role played by entanglement during such dissipative quantum evolution. We have shown that it is the entanglement of formation the metric to be used as the  emmiters' quantifier  of entanglement and that the use of concurrence for this purpose can give misleading results about the actual degree of generated entanglement. 
We have also shown that the dynamics of correlations followed by the emitters can be controlled and engineered in the presence of a coherent laser field. We have quantified and illustrated the interplay between entanglement and quantum discord stressing the importance of the persistent quantum correlations in several different scenarios where entanglement vanishes.

 In cases where 
classical correlations are greater than quantum correlations,
a coherent control mechanism can  be implemented so that  the latter are enhanced and  become greater than the former.  This result is corroborated for driven emitters in the case of strong  laser-dipole coupling: by preparing the initial symmetric state, once entanglement vanishes there is a rapid stabilisation of correlations and the discord becomes  greater  than the  classical correlations. Moreover, for strong laser excitation, now starting with the antisymmetric state, the discord is, at all times,  greater  than the  classical correlations, regardless of the emitters' entanglement content. This  result is resilient to scenarios where the collective (incoherent) decay rate reflects a strong coupling of the emitters to the surrounding environment. In fact, the strength of the quantum correlations can profit from  large collective  decay rates, if the emitters strength coupling and the laser field excitation are appropriately tuned.

From an experimental viewpoint,  set-ups comprising single molecules \cite{Hettich,niek1} or semiconductor quantum dots \cite{ander} could be employed to demonstrate our findings. In particular,  single-molecule techniques that allow femtosecond pulse-shaping are currently  available, and controlled coherent superpositions and  basic femtosecond single-qubit operations in single organic molecules have been produced and manipulated even at room temperature \cite{niek1}. 

\section{Acknowledgements}

We gratefully acknowledge Universidad del Valle Vice Presidency of Research under contract CI 7859. We acknowledge R. Hildner and N. F. van Hulst for discussions.

\end{document}